# Fundamentals and applications of Van der Waals magnets in magnon spintronics

Samuel Mañas-Valero,[1] Toeno van der Sar,[1] Rembert A. Duine,[2,3] and Bart van Wees[4]

[1]Department of Quantum Nanoscience, Kavli Institute of Nanoscience, Delft University of Technology, Delft 2628CJ, The Netherlands
[2]Institute for Theoretical Physics, Utrecht University, 3584CC Utrecht, The Netherlands
[3]Department of Applied Physics, Eindhoven University of Technology, 5612 AP Eindhoven, The Netherlands
[4]Zernike Institute for Advanced Materials, University of Groningen, NL-9747AG Groningen, The Netherlands
*Correspondence: S.ManasValero@tudelft.nl

## SUMMARY

Spintronics is concerned with replacing charge current with current of spin, the electron's intrinsic angular momentum. In magnetic insulators, spin currents are carried by magnons, the quanta of spin-wave excitations on top of the magnetically ordered state. Magnon spin currents are especially promising for information technology due to their low intrinsic damping, non-reciprocal transport, micrometer wavelengths at microwave frequencies, and strong interactions that enable signal transduction. In this perspective, we give our view on the progress and challenges towards realizing magnon spintronics based on atomically thin Van der Waals magnets, a recently discovered class of magnetic materials of which the tunability and versatility has attracted a great deal of ongoing research.

## INTRODUCTION

Many-particle systems tend to order at low temperatures. Often, the order gives rise to emergent collective modes that correspond to wavelike small-amplitude fluctuations around the ordered state that do not exist in the disordered state. A prime example from solid-state physics is the solid itself. Its atoms are ordered on a periodic lattice structure, giving rise to lattice vibrations –of which the quanta are dubbed phonons– that carry energy, quasi-momentum and even spin angular momentum.[1] Another type of collective modes, that carry spin angular momentum, are spin waves of which the quanta are dubbed magnons. These are the elementary excitations on top of magnetically ordered states. The simplest example of a spin wave or magnon is to consider a ferromagnetically-ordered system of spin one-half particles that are coupled ferromagnetically. Flipping one of the spins creates an excitation that transports one quantum of angular momentum opposite to the spin direction of the ferromagnetic order. Due to the coupling between the spins, this localized spin flip is not an eigenstate but delocalizes into a wavelike excitation **(Figure 1a)**.

Spin waves are sometimes referred to as the Goldstone boson that emerges from the spontaneously broken continuous spin-rotation symmetry. This language may, however, not be completely adequate. In magnetic materials there are anisotropies that favor certain directions of magnetization. Therefore, the magnetically ordered state in practice never spontaneously breaks a continuous symmetry and the spin-wave spectrum is gapped with a gap equal to the ferromagnetic resonance frequency. The symmetry that all magnetically ordered states break is time-reversal symmetry, a discrete symmetry. Moreover, the word *"boson"* suggests that the magnon is a boson simply because its angular momentum is an integer (namely one) times the quantum of angular momentum. Magnons are, however, bosonic because they are in linear approximation described as the quasiparticle excitations of quantum harmonic oscillators, like photons and phonons. In the above example of the chain of spin-one-half particles, the magnon is, for example, a scalar, *i.e.*, a spinless, bosonic excitation that transports only one component of the spin angular momentum, whereas a true spinful particle would be able to transport all vector components. As another example, a magnon on top of an antiferromagnetically ordered state is an effective two-level boson with one polarization carrying a positive quantum of angular momentum whereas the other polarization carries a negative quantum of angular momentum.[2]

Magnons appear in various guises, ranging from the simple example of a spin wave on top of a ferromagnetically-ordered state (**Figure 1**), to complicated excitations on non-collinear magnetic ground states that may break discrete and/or continuous symmetries that are very different from continuous spin-rotation symmetry.[3] An example is the skyrmion lattice (a lattice of microscopic, whirlpool-like magnetization textures) that breaks time-reversal and continuous-translation symmetry and has several unique collective excitations. This points to the attractive feature that magnons can be excited, detected and manipulated by external control knobs in various ways (**Figure 1**). One is to apply external stimuli, such as an external field, to control the properties of magnons –for example, the gap in their spectrum– directly. Another is to influence the properties of magnons by controlling the magnetically ordered state on top of which they are the excitations. This makes possible that different types of magnons emerge in one material and to reversibly go from one type to another.

The field that seeks to utilize the properties of magnons is dubbed *"magnonics"* or *"magnon spintronics"*.[4] Other fields that have the goal of utilizing collective excitations, such as *"phononics"*, *"nanophotonics"*, and *"plasmonics"* all seek to manipulate these excitations in one way or another and each have their own benefits. The flexibility to use different magnetically ordered states in one material to endow spin waves with different properties is, however, unique to magnon spintronics. Moreover, the wavelike nature of magnons so that they carry both amplitude and phase, their strongly confined magnetic stray fields, their frequency range that goes from sub GHz to THz, their low damping, the non-reciprocal effects, the strong non-linearities, absence of Joule heating, and resulting long propagation length, and the ability to operate both in the classical and quantum regime,[5] further underly the potential of this nascent field for applications.

So far, magnon spintronics has used yttrium-iron-garnet (YIG) and a few other materials as the main workhorses to achieve many of its recent results. These include demonstration of logic devices and transistors with coherent spin waves,[6] conversion of thermal magnon spin currents into electronic spin current via the spin Seebeck effect,[7] and long-distance propagation of spin currents through ferromagnets and antiferromagnets.[8,9] The discovery of ultrathin Van der Waals magnets with stable and tunable properties[10] has opened up a new perspective for magnon spintronics. The low dimensionality of the system together with the ability to tune the magnetic properties, such as exchange and anisotropy by gating, and the possibility to combine and twist different materials forming van der Waals heterostructures, gives a rich and novel playground for magnon spintronics.[11,12]

Van der Waals magnets arise either naturally –*i.e.*, in layered magnets that can be exfoliated down to the atomic limit– or by design –that is, by engineering proximity or twist effects in Van der Waals heterostructures–, even if the starting layers are not magnetic *per se*, as in the case of graphene.[13,14] In addition, beyond the large and diverse nature of Van der Waals magnets, the know-how on the manipulation of two-dimensional materials developed in the last 20 years brings new tuning knobs with easy experimental implementation that are absent in conventional magnonic materials as, for instance, electrical gating,[15] strain,[16] stacking[17,18] or twisting[19] engineering, therefore offering a roadmap for overcoming current experimental challenges in conventional magnon spintronics. For example, proposals for spin superfluidity,[20] topologically protected magnon spin currents,[21] and magnon spin transistors,[6,22] may ultimately be realized using Van der Waals magnets. In this perspective, we give our view on the challenges, progress, and perspective of developing magnon spintronics with Van der Waals magnets.

## GENERATION AND DECTION OF MAGNONS IN VAN DER WAALS MAGNETS

Magnon spintronics typically relies on two ways of exciting and/or detecting magnons. These are coherent excitation and detection with radiofrequency (RF) antennas[4] and incoherent excitation by injection or extraction of spin current from a heavy-metal reservoir that utilizes the spin Hall effect or its inverse to convert the electronic spin currents into charge currents or voltages.[7,8,13]

### *Generation and detection of coherent spin waves*

In the prototypical coherent spin-wave transport experiment, one antenna excites a spin wave of a certain frequency and wave vector and a second antenna at some distance picks up the spin wave inductively. Such inductive methods are however challenging to apply in nanometer-thin Van der Waals magnets because of the lack of magnetic volume and resulting small magnetic signals.[23] In addition, the spin-wave damping in Van der Waals magnets is often unknown, hampering predictions of the expected spin-wave amplitudes. While this prototypical experiment has not been demonstrated on Van der Waals magnets yet, many experiments have now reported the inductive excitation and detection of the zero-wave-vector modes typically referred to as the ferromagnetic resonance (FMR). For instance, MacNeill *et al.* demonstrated FMR detection in $CrCl_3$,[24] In this material, ferromagnetically ordered atomic planes couple antiferromagnetically. Because the inter-layer magnetic coupling is weak, the resulting antiferromagnetic resonances are in the GHz regime that is well suited for inductive methods. Similar results have been obtained on $CrPS_4$, CrSBr, $CrCl_3$, and on Van der Waals ferromagnets such as $Cr_2Ge_2Te_6$ or $Fe_5GeTe_2$, among others.[10,23]

The central challenge for probing coherent spin wave transport in Van der Waals magnets using inductive methods is the small magnetic fields generated by these atomically thin materials.[10,23] As such, high-sensitivity local-probe techniques offer a promising alternative path for coherent spin wave imaging. X-ray microscopy recently demonstrated signatures of coherent, finite-wavelength spin wave propagation in 30-nm thick $Fe_5GeTe_2$.[25] Alternatively, nitrogen-vacancy (NV) spins in diamond provide high-resolution imaging of static and dynamic magnetic stray fields. Motivated by successes in imaging magnetizations of few-layer Van der Waals magnets,[26,27] a growing effort is pushing NV microscopy towards imaging coherent spin wave transport in these materials. While this goal has not been reached yet, recent NV experiments have reported the detection of incoherent spin dynamics in thin Van der Waals magnets.[28,29] A challenge for NV-based coherent spin wave imaging is that the detection frequency is limited by the electron spin resonance frequencies of the NV sensor spin.

In addition to microwave resonance experiments, magnons in Van der Waals magnets have been detected via their coupling to other degrees of freedom, such as phonons or polarons, and via spectroscopy methods based on diverse frequency ranges of the electromagnetic field.[10,23,30] Because many Van der Waals magnets are also semiconductors, strong spin-exciton coupling has enabled coherent excitation and detection of magnon wave packets.[31] Light also enables controlling the magnetic anisotropy and thereby the magnons with ultrashort pulses of light, as demonstrated in $NiPS_3$.[32] Similarly, in two dimensions the magnon and plasmon dispersions may cross which enables strong magnon-plasmon coupling, as theoretically discussed by Ghosh *et al.*[33] With a few exceptions,[23] the generation, control, and detection of coherent spin-wave transport -of particular interest for creating devices- in atomically thin Van der Waals magnets has not been demonstrated yet. Only recently, the antiferromagnetic resonance in spin-filter tunnelling junctions based on $PtTe_2$/bilayer CrSBr/graphite has been detected,[34] thus bridging the gap between direct currents and high-frequency spin dynamics.

### *Injection and detection of thermal magnons*

So far, we have discussed the progress towards and challenges for injection and detection of spin waves at specific frequencies, *i.e.*, in the coherent regime. Due to thermal fluctuations, magnons are present at all energies with an occupation given by the Bose-Einstein distribution. This thermal magnon gas can be biased by thermal gradients and/or injection and extraction of spin current from heavy-metal leads. Due to the strong spin-orbit coupling in the latter, a charge current tangential to the interface between the heavy metal and the magnet leads –via the spin Hall effect– to an electronic spin current that impinges on the interface between metal and magnet and is converted into a magnon spin current via interfacial exchange interactions. The strength of this interaction may be probed from the spin Hall magnetoresistance, as demonstrated with various Van der Waals magnets where the magnetic order is probed electrically.[10]

The conversion of electron to magnon spin currents was first probed in the spin Seebeck effect in which a magnon spin current generated by a thermal gradient is detected as an electronic charge current via the inverse spin Hall effect.[7] Cornelissen *et al.* demonstrated long-range magnon spin currents through YIG via injection and detection with Pt leads.[22] Similar experiments were conducted by Lebrun *et al.* on antiferromagnetic hematite.[9] In these types of experiments, thermally-generated magnon spin currents are distinguished from electrically-generated once via lock-in techniques: an alternating current with low frequency is passed through the heavy-metal injector. The first-harmonic signal in the heavy-metal detector is the result of injection of spin from the injector, whereas the second-harmonic signal in the detector stems from the Joule heating of the injector that sets up a temperature gradient that drives the magnons.

Coupling between electron spins and magnons in Van der Waals magnets was demonstrated early on in magnon-assisted tunneling between graphene layers through $CrBr_3$.[35] The first experiments with heavy-metal leads interfaced with Van der Waals magnets showed evidence for thermal magnon currents, developing Van der Waals magnon valves.[36] As an example of the tunability of the magnon transport, Feringa *et al.* detected the spin-flop transition via these thermal magnon currents.[37] The challenge to overcome to observe electrical spin injection rather than these thermally generated magnon currents is to increase the interfacial exchange interaction at the interface on which the spin injection relies. Recently, De Wal *et al.*[38] have demonstrated spin injection and detection in $CrPS_4$. In the same material, Qi *et al.*[39] were able to show a large modulation of thermal magnon spin currents. In this regard, the difference in transport between thermally generated magnons (via the spin Seebeck effect) and electrically injected magnons (via the spin Hall effect) has been studied by De Wal *et al.*[40] in the non-local geometry (see **Figure 2**), showing that, due to the magnetic field dependence of the specific magnon modes in the antiferromagnet $CrPS_4$, electrically generated magnon transport could only be observed above the spin-flip magnetic field of about 7 T. In contrast, the non-locally observed spin Seebeck effect exhibits a much more complicated dependence on magnetic field. Finally, a first demonstration of magnon spin transistors in Van der Waals magnets was given, where the magnon transport was modulated by thermal or electrical magnon injection from a gate electrode.[41]

## OUTLOOK

We have discussed recent rapid progress regarding detection and manipulation of magnons with the long-term perspective of realizing magnon spintronics with Van der Waals magnets. Despite this progress, many challenges remain: for many magnon spintronics applications, one would need a Van der Waals magnet with a critical temperature well above room temperature, with strong interfacial magnon-electron coupling with a heavy metal, and with tunable magnetic properties. The first of these challenges is not specific to magnon spintronics but to all room-temperature applications of Van der Waals magnets. Fortunately, there are now Van der Waals magnets with critical temperature above room temperature, as $Fe_5GeTe_3$ and related compounds,[10] where the ordering temperature can be tuned based on doping, representing a versatile chemical route for tuning their properties *a la carte* while preserving their magnetic properties down to the atomically thin limit. Indeed, many materials and material combination are currently being explored.[10,23] The existence of spin Hall magnetoresistive effects and electrically actuated spin transport through some Van der Waals materials show that there are material combinations that enable interfacial spin transport. Finally, progress has been made towards controlling this spin transport via external fields or electrical modulators.

Many of the experiments that we discussed so far have been carried out on relatively thick samples because the strength of the various signals typically scales with sample thickness. It remains a challenge to inject, control, and detect magnons in atomically thin samples. In ultrathin YIG, a very large magnon spin conductivity was demonstrated in the ultrathin limit that is not fully understood.[42] This result is, however, a strong motivation to go to atomically thin Van der Waals magnets and probe their magnon transport. Moreover, such ultrathin samples would allow for efficient manipulation of their magnetic properties, *e.g.*, via transistor-like gating and injection of spin current from a modulator. For large spin currents, such injection may produce Bose-Einstein-condensation-like instabilities that could give rise to spin superfluid transport.[43] Spin superfluid transport is also expected in magnets with easy-plane anisotropy,[20] such as in an uniaxial antiferromagnet with field larger than the spin flop field.[44] Here, the angle in the easy plane is the phase of the superfluid whose winding carries the superfluid current. Van der Waals magnets with such easy-plane anisotropy, such as $CrCl_3$, have been found.[10] The challenge is now to inject spin current with spin polarization perpendicular to the easy plane, which could be done using the spin Hall effect in ferromagnets.[45] Despite promising results,[46] electrically actuated superfluid spin transport –with its unique signatures of upper and lower critical current and drop in thermal spin current because it is short-circuited by the superfluid–[20,47] has not been conclusively demonstrated. Nonetheless, Van der Waals magnets appear to be the ideal platform to push this direction further, eventually perhaps enabling such superfluid spin transport even at room temperature.

Van der Waals magnets also appear to be the ideal platform to demonstrate spin transport via topologically protected magnon edge states.[21] These have been predicted to exist in several Van der Waals magnets. While magnon bands have been observed in bulk crystals via neutron scattering,[48] their topology has not been directly demonstrated. For coherent excitation of the topological edge states, the challenge is that they exist at large frequency. It was recently theoretically proposed, however, that in an out-of-equilibrium situation the topologically protected edge states may be brought down to zero frequency,[49] thereby enabling their detection via RF techniques. For incoherent excitation of topologically protected magnon transport, the challenge is to distinguish the contribution of the bulk from the edge. It was shown theoretically that in principle the edge and bulk give rise to two different length scales which would allow one to distinguish them.[50] Very recently, edge states were detected using scanning-tunneling microscopy on atomically thin $CrI_3$ flakes.[51]

Generically speaking, going to the true two-dimensional limit opens a large vista for efficient manipulation and interesting physics due to the increasing importance of interactions and fluctuations. A very recent example is the evidence for hydrodynamic behavior of magnons[29] which is the result of strong magnon-magnon interactions that gives rise magnon viscous effects that may also be detected in the magnon spin transport.[52] Another example is the possibility to manipulate magnon properties via twisted stacking of two Van der Waals magnets.[19] Thus, we expect Van der Waals magnets in magnon spintronics to be an expanding area of research in the near future where determining the quantum metrics for magnon transport -as the damping, coherence length, band topology or magnonic heat capacity-[4] will be the first step towards the realization of potential emergent applications in areas such as neuromorphic computing,[53] or quantum magnonics[5] and quantum information systems.[54,55] As a second step, we envision 2D magnets as a platform for coupling magnons with other quasiparticles, as phonons, plasmons or photons, unraveling new dynamic coupling mechanisms for realizing the ultimate strong coupling regime when interfaced, for example, with optical cavities or superconducting resonators.[56,57]In view of the wide range of possibilities and speed of recent progress, we are convinced that many spectacular results on magnon spintronics with Van der Waals magnets will be demonstrated in the years to come.

***Lead contact***

Further information and requests for resources should be directed to and will be fulfilled by the lead contact, Samuel Mañas-Valero (S.ManasValero@tudelft.nl).

## ACKNOWLEDGMENTS

This work was supported by the Dutch Research Council (NWO) via grant OCENW.XL21.XL21.058. S.M.-V. acknowledges the support from the European Commission for a Marie Sklodowska–Curie individual fellowship

No. 101103355-SPIN-2D-LIGHT. BvW acknowledges support from the European Research Council (ERC) under the European Union's 2DMAGSPIN (Grant Agreement No. 101053054).

## DECLARATION OF INTERESTS

R. A. is an advisory board member for Newton. The authors declare no other competing interests.

## DECLARATION OF GENERATIVE AI AND AI-ASSISTED TECHNOLOGIES

No generative or AI-assisted technologies were employed in this work.

## REFERENCES

1. Holanda, J., Maior, D.S., Azevedo, A., and Rezende, S.M. (2018). Detecting the phonon spin in magnon-phonon conversion experiments. Nat Phys *14*, 500–506. https://doi.org/10.1038/s41567-018-0079-y.

2. Gückelhorn, J., de-la-Peña, S., Scheufele, M., Grammer, M., Opel, M., Geprägs, S., Cuevas, J.C., Gross, R., Huebl, H., Kamra, A., et al. (2023). Observation of the Nonreciprocal Magnon Hanle Effect. Phys Rev Lett *130*, 216703. https://doi.org/10.1103/PhysRevLett.130.216703.

3. Yu, H., Xiao, J., and Schultheiss, H. (2021). Magnetic texture based magnonics. Phys Rep *905*, 1–59. https://doi.org/10.1016/j.physrep.2020.12.004.

4. Chumak, A. V., Vasyuchka, V.I., Serga, A.A., and Hillebrands, B. (2015). Magnon spintronics. Nat Phys *11*, 453–461. https://doi.org/10.1038/nphys3347.

5. Yuan, H.Y., Cao, Y., Kamra, A., Duine, R.A., and Yan, P. (2022). Quantum magnonics: When magnon spintronics meets quantum information science. Phys Rep *965*, 1–74. https://doi.org/10.1016/j.physrep.2022.03.002.

6. Chumak, A. V., Serga, A.A., and Hillebrands, B. (2014). Magnon transistor for all-magnon data processing. Nat Commun *5*, 4700. https://doi.org/10.1038/ncomms5700.

7. Uchida, K., Xiao, J., Adachi, H., Ohe, J., Takahashi, S., Ieda, J., Ota, T., Kajiwara, Y., Umezawa, H., Kawai, H., et al. (2010). Spin Seebeck insulator. Nat Mater *9*, 894–897. https://doi.org/10.1038/nmat2856.

8. Cornelissen, L.J., Liu, J., Duine, R.A., Youssef, J. Ben, and van Wees, B.J. (2015). Long-distance transport of magnon spin information in a magnetic insulator at room temperature. Nat Phys *11*, 1022–1026. https://doi.org/10.1038/nphys3465.

9. Lebrun, R., Ross, A., Bender, S.A., Qaiumzadeh, A., Baldrati, L., Cramer, J., Brataas, A., Duine, R.A., and Kläui, M. (2018). Tunable long-distance spin transport in a crystalline antiferromagnetic iron oxide. Nature *561*, 222–225. https://doi.org/10.1038/s41586-018-0490-7.

10. Wang, Q.H., Bedoya-Pinto, A., Blei, M., Dismukes, A.H., Hamo, A., Jenkins, S., Koperski, M., Liu, Y., Sun, Q.-C., Telford, E.J., et al. (2022). The Magnetic Genome of Two-Dimensional van der Waals Materials. ACS Nano *16*, 6960–7079. https://doi.org/10.1021/acsnano.1c09150.

11. Jiang, S., Li, L., Wang, Z., Mak, K.F., and Shan, J. (2018). Controlling magnetism in 2D CrI3 by electrostatic doping. Nat Nanotechnol *13*, 549–553. https://doi.org/10.1038/s41565-018-0135-x.

12. Boix-Constant, C., Jenkins, S., Rama-Eiroa, R., Santos, E.J.G., Mañas-Valero, S., and Coronado, E. (2024). Multistep magnetization switching in orthogonally twisted ferromagnetic monolayers. Nat Mater *23*, 212–218. https://doi.org/10.1038/s41563-023-01735-6.

13. Wei, D.S., van der Sar, T., Lee, S.H., Watanabe, K., Taniguchi, T., Halperin, B.I., and Yacoby, A. (2018). Electrical generation and detection of spin waves in a quantum Hall ferromagnet. Science (1979) *362*, 229–233. https://doi.org/10.1126/science.aar4061.

14. Lu, X., Stepanov, P., Yang, W., Xie, M., Aamir, M.A., Das, I., Urgell, C., Watanabe, K., Taniguchi, T., Zhang, G., et al. (2019). Superconductors, orbital magnets and correlated states in magic-angle bilayer graphene. Nature *574*, 653–657. https://doi.org/10.1038/s41586-019-1695-0.

15. Wu, F., Gibertini, M., Watanabe, K., Taniguchi, T., Gutiérrez-Lezama, I., Ubrig, N., and Morpurgo, A.F. (2023). Gate-Controlled Magnetotransport and Electrostatic Modulation of Magnetism in 2D Magnetic Semiconductor $CrPS_4$. Advanced Materials *35*. https://doi.org/10.1002/adma.202211653.

16. Esteras, D.L., Rybakov, A., Ruiz, A.M., and Baldoví, J.J. (2022). Magnon Straintronics in the 2D van der Waals Ferromagnet CrSBr from First-Principles. Nano Lett *22*, 8771–8778. https://doi.org/10.1021/acs.nanolett.2c02863.

17. Wang, C., Gao, Y., Lv, H., Xu, X., and Xiao, D. (2020). Stacking Domain Wall Magnons in Twisted van der Waals Magnets. Phys Rev Lett *125*, 247201. https://doi.org/10.1103/PhysRevLett.125.247201.

18. Boix-Constant, C., Rybakov, A., Miranda-Pérez, C., Martínez-Carracedo, G., Ferrer, J., Mañas-Valero, S., and Coronado, E. (2025). Programmable Magnetic Hysteresis in Orthogonally-Twisted 2D CrSBr Magnets via Stacking Engineering. Advanced Materials. https://doi.org/10.1002/adma.202415774.

19. Li, Y.-H., and Cheng, R. (2020). Moiré magnons in twisted bilayer magnets with collinear order. Phys Rev B *102*, 094404. https://doi.org/10.1103/PhysRevB.102.094404.

20. Sonin, E.B. (2010). Spin currents and spin superfluidity. Adv Phys *59*, 181–255. https://doi.org/10.1080/00018731003739943.

21. Owerre, S.A. (2016). A first theoretical realization of honeycomb topological magnon insulator. Journal of Physics: Condensed Matter *28*, 386001. https://doi.org/10.1088/0953-8984/28/38/386001.

22. Cornelissen, L.J., Liu, J., van Wees, B.J., and Duine, R.A. (2018). Spin-Current-Controlled Modulation of the Magnon Spin Conductance in a Three-Terminal Magnon Transistor. Phys Rev Lett *120*, 097702. https://doi.org/10.1103/PhysRevLett.120.097702.

23. Tang, C., Alahmed, L., Mahdi, M., Xiong, Y., Inman, J., McLaughlin, N.J., Zollitsch, C., Kim, T.H., Du, C.R., Kurebayashi, H., et al. (2023). Spin dynamics in van der Waals magnetic systems. Phys Rep *1032*, 1–36. https://doi.org/10.1016/j.physrep.2023.09.002.

24. MacNeill, D., Hou, J.T., Klein, D.R., Zhang, P., Jarillo-Herrero, P., and Liu, L. (2019). Gigahertz Frequency Antiferromagnetic Resonance and Strong Magnon-Magnon Coupling in the Layered Crystal $CrCl_3$. Phys Rev Lett *123*, 047204. https://doi.org/10.1103/PhysRevLett.123.047204.

25. Schulz, F., Litzius, K., Powalla, L., Birch, M.T., Gallardo, R.A., Satheesh, S., Weigand, M., Scholz, T., Lotsch, B. V., Schütz, G., et al. (2023). Direct Observation of Propagating Spin Waves in the 2D van der Waals Ferromagnet $Fe_5GeTe_2$. Nano Lett *23*, 10126–10131. https://doi.org/10.1021/acs.nanolett.3c02212.

26. Thiel, L., Wang, Z., Tschudin, M.A., Rohner, D., Gutiérrez-Lezama, I., Ubrig, N., Gibertini, M., Giannini, E., Morpurgo, A.F., and Maletinsky, P. (2019). Probing magnetism in 2D materials at the nanoscale with single-spin microscopy. Science *364*, 973–976. https://doi.org/10.1126/science.aav6926.

27. Ghiasi, T.S., Borst, M., Kurdi, S., Simon, B.G., Bertelli, I., Boix-Constant, C., Mañas-Valero, S., van der Zant, H.S.J., and van der Sar, T. (2023). Nitrogen-vacancy magnetometry of CrSBr by diamond membrane transfer. NPJ 2D Mater Appl *7*, 62. https://doi.org/10.1038/s41699-023-00423-y.

28. Huang, M., Green, J.C., Zhou, J., Williams, V., Li, S., Lu, H., Djugba, D., Wang, H., Flebus, B., Ni, N., et al. (2023). Layer-Dependent Magnetism and Spin Fluctuations in Atomically Thin van der Waals Magnet $CrPS_4$. Nano Lett *23*, 8099–8105. https://doi.org/10.1021/acs.nanolett.3c02129.

29. Xue, R., Maksimovic, N., Dolgirev, P.E., Xia, L.-Q., Kitagawa, R., Müller, A., Machado, F., Klein, D.R., MacNeill, D., Watanabe, K., et al. (2024). Signatures of magnon hydrodynamics in an atomically-thin ferromagnet. Arxiv: 2403.01057.

30. Lyons, T.P., Puebla, J., Yamamoto, K., Deacon, R.S., Hwang, Y., Ishibashi, K., Maekawa, S., and Otani, Y. (2023). Acoustically Driven Magnon-Phonon Coupling in a Layered Antiferromagnet. Phys Rev Lett *131*, 196701. https://doi.org/10.1103/PhysRevLett.131.196701.

31. Bae, Y.J., Wang, J., Scheie, A., Xu, J., Chica, D.G., Diederich, G.M., Cenker, J., Ziebel, M.E., Bai, Y., Ren, H., et al. (2022). Exciton-coupled coherent magnons in a 2D semiconductor. Nature *609*, 282–286. https://doi.org/10.1038/s41586-022-05024-1.

32. Afanasiev, D., Hortensius, J.R., Matthiesen, M., Mañas-Valero, S., Šiškins, M., Lee, M., Lesne, E., van der Zant, H.S.J., Steeneken, P.G., Ivanov, B.A., et al. (2021). Controlling the anisotropy of a van der Waals antiferromagnet with light. Sci Adv *7*, eabf3096. https://doi.org/10.1126/sciadv.abf3096.

33. Ghosh, S., Menichetti, G., Katsnelson, M.I., and Polini, M. (2023). Plasmon-magnon interactions in two-dimensional honeycomb magnets. Phys Rev B *107*, 195302. https://doi.org/10.1103/PhysRevB.107.195302.

34. Cham, T.M.J., Chica, D.G., Watanabe, K., Taniguchi, T., Roy, X., Luo, Y.K., and Ralph, D.C. (2024). Spin-filter tunneling detection of antiferromagnetic resonance with electrically-tunable damping. Arxiv 2407.09462.

35. Ghazaryan, D., Greenaway, M.T., Wang, Z., Guarochico-Moreira, V.H., Vera-Marun, I.J., Yin, J., Liao, Y., Morozov, S. V, Kristanovski, O., Lichtenstein, A.I., et al. (2018). Magnon-assisted tunnelling in van der Waals heterostructures based on $CrBr_3$. Nat Electron *1*, 344–349. https://doi.org/10.1038/s41928-018-0087-z.

36. Chen, G., Qi, S., Liu, J., Chen, D., Wang, J., Yan, S., Zhang, Y., Cao, S., Lu, M., Tian, S., et al. (2021). Electrically switchable van der Waals magnon valves. Nat Commun *12*, 6279. https://doi.org/10.1038/s41467-021-26523-1.

37. Feringa, F., Vink, J.M., and van Wees, B.J. (2022). Spin-flop transition in the quasi-two-dimensional antiferromagnet $MnPS_3$ detected via thermally generated magnon transport. Phys Rev B *106*, 224409. https://doi.org/10.1103/PhysRevB.106.224409.

38. de Wal, D.K., Iwens, A., Liu, T., Tang, P., Bauer, G.E.W., and van Wees, B.J. (2023). Long-distance magnon transport in the van der Waals antiferromagnet $CrPS_4$. Phys Rev B *107*, L180403. https://doi.org/10.1103/PhysRevB.107.L180403.

39. Qi, S., Chen, D., Chen, K., Liu, J., Chen, G., Luo, B., Cui, H., Jia, L., Li, J., Huang, M., et al. (2023). Giant electrically tunable magnon transport anisotropy in a van der Waals antiferromagnetic insulator. Nat Commun *14*, 2526. https://doi.org/10.1038/s41467-023-38172-7.

40. de Wal, D.K., Zohaib, M., and van Wees, B.J. (2024). Magnon spin transport in the van der Waals antiferromagnet <math> <msub> <mi>CrPS</mi> <mn>4</mn> </msub> </math> for noncollinear and collinear magnetization. Phys Rev B *110*, 174440. https://doi.org/10.1103/PhysRevB.110.174440.

41. de Wal, D.K., Mena, R.L., Zohaib, M., and van Wees, B.J. (2024). Gate control of magnon spin transport in unconventional magnon transistors based on the van der Waals antiferromagnet <math> <msub> <mrow> <mi>CrPS</mi> </mrow> <mn>4</mn> </msub> </math>. Phys Rev B *110*, 224434. https://doi.org/10.1103/PhysRevB.110.224434.

42. Wei, X.Y., Santos, O.A., Lusero, C.H.S., Bauer, G.E.W., Ben Youssef, J., and van Wees, B.J. (2022). Giant magnon spin conductivity in ultrathin yttrium iron garnet films. Nat Mater *21*, 1352–1356. https://doi.org/10.1038/s41563-022-01369-0.

43. Bender, S.A., Duine, R.A., and Tserkovnyak, Y. (2012). Electronic Pumping of Quasiequilibrium Bose-Einstein-Condensed Magnons. Phys Rev Lett *108*, 246601. https://doi.org/10.1103/PhysRevLett.108.246601.

44. Takei, S., and Tserkovnyak, Y. (2014). Superfluid Spin Transport Through Easy-Plane Ferromagnetic Insulators. Phys Rev Lett *112*, 227201. https://doi.org/10.1103/PhysRevLett.112.227201.

45. Das, K.S., Schoemaker, W.Y., van Wees, B.J., and Vera-Marun, I.J. (2017). Spin injection and detection via the anomalous spin Hall effect of a ferromagnetic metal. Phys Rev B *96*, 220408. https://doi.org/10.1103/PhysRevB.96.220408.

46. Yuan, W., Zhu, Q., Su, T., Yao, Y., Xing, W., Chen, Y., Ma, Y., Lin, X., Shi, J., Shindou, R., et al. (2018). Experimental signatures of spin superfluid ground state in canted antiferromagnet $Cr_2O_3$ via nonlocal spin transport. Sci Adv *4*. https://doi.org/10.1126/sciadv.aat1098.

47. Flebus, B., Bender, S.A., Tserkovnyak, Y., and Duine, R.A. (2016). Two-Fluid Theory for Spin Superfluidity in Magnetic Insulators. Phys Rev Lett *116*, 117201. https://doi.org/10.1103/PhysRevLett.116.117201.

48. Chen, L., Chung, J.-H., Gao, B., Chen, T., Stone, M.B., Kolesnikov, A.I., Huang, Q., and Dai, P. (2018). Topological Spin Excitations in Honeycomb Ferromagnet $CrI_3$. Phys Rev X *8*, 041028. https://doi.org/10.1103/PhysRevX.8.041028.

49. Gunnink, P.M., Harms, J.S., Duine, R.A., and Mook, A. (2023). Zero-Frequency Chiral Magnonic Edge States Protected by Nonequilibrium Topology. Phys Rev Lett *131*, 126601. https://doi.org/10.1103/PhysRevLett.131.126601.

50. Rückriegel, A., Brataas, A., and Duine, R.A. (2018). Bulk and edge spin transport in topological magnon insulators. Phys Rev B *97*, 081106. https://doi.org/10.1103/PhysRevB.97.081106.

51. Zhang, J., Zhang, M.-H., Li, P., Liu, Z., Tao, Y., Wang, H., Yao, D.-X., Guo, D., and Zhong, D. (2024). Direct observation of topological magnon edge states. Arxiv:2410.18960.

52. Ulloa, C., Tomadin, A., Shan, J., Polini, M., van Wees, B.J., and Duine, R.A. (2019). Nonlocal Spin Transport as a Probe of Viscous Magnon Fluids. Phys Rev Lett *123*, 117203. https://doi.org/10.1103/PhysRevLett.123.117203.

53. Menezes, R.M., Mulkers, J., de Souza Silva, C.C., Van Waeyenberge, B., and Milošević, M. V. (2024). Toward magnonic logic and neuromorphic computing: controlling spin waves by spin-polarized current. Phys Rev Appl *22*, 054056. https://doi.org/10.1103/PhysRevApplied.22.054056.

54. Pal, P.K., Mondal, A.K., and Barman, A. (2024). Using magnons as a quantum technology platform: a perspective. Journal of Physics Condensed Matter *36*. https://doi.org/10.1088/1361-648X/ad6828.

55. Pirro, P., Vasyuchka, V.I., Serga, A.A., and Hillebrands, B. (2021). Advances in coherent magnonics. Preprint at Nature Research, https://doi.org/10.1038/s41578-021-00332-w https://doi.org/10.1038/s41578-021-00332-w.

56. Bozhko, D.A., Vasyuchka, V.I., Chumak, A. V., and Serga, A.A. (2020). Magnon-phonon interactions in magnon spintronics (Review article). Low Temperature Physics *46*, 383–399. https://doi.org/10.1063/10.0000872.

57. Zollitsch, C.W., Khan, S., Nam, V.T.T., Verzhbitskiy, I.A., Sagkovits, D., O'Sullivan, J., Kennedy, O.W., Strungaru, M., Santos, E.J.G., Morton, J.J.L., et al. (2023). Probing spin dynamics of ultra-thin van der Waals magnets via photon-magnon coupling. Nat Commun *14*, 2619. https://doi.org/10.1038/s41467-023-38322-x.

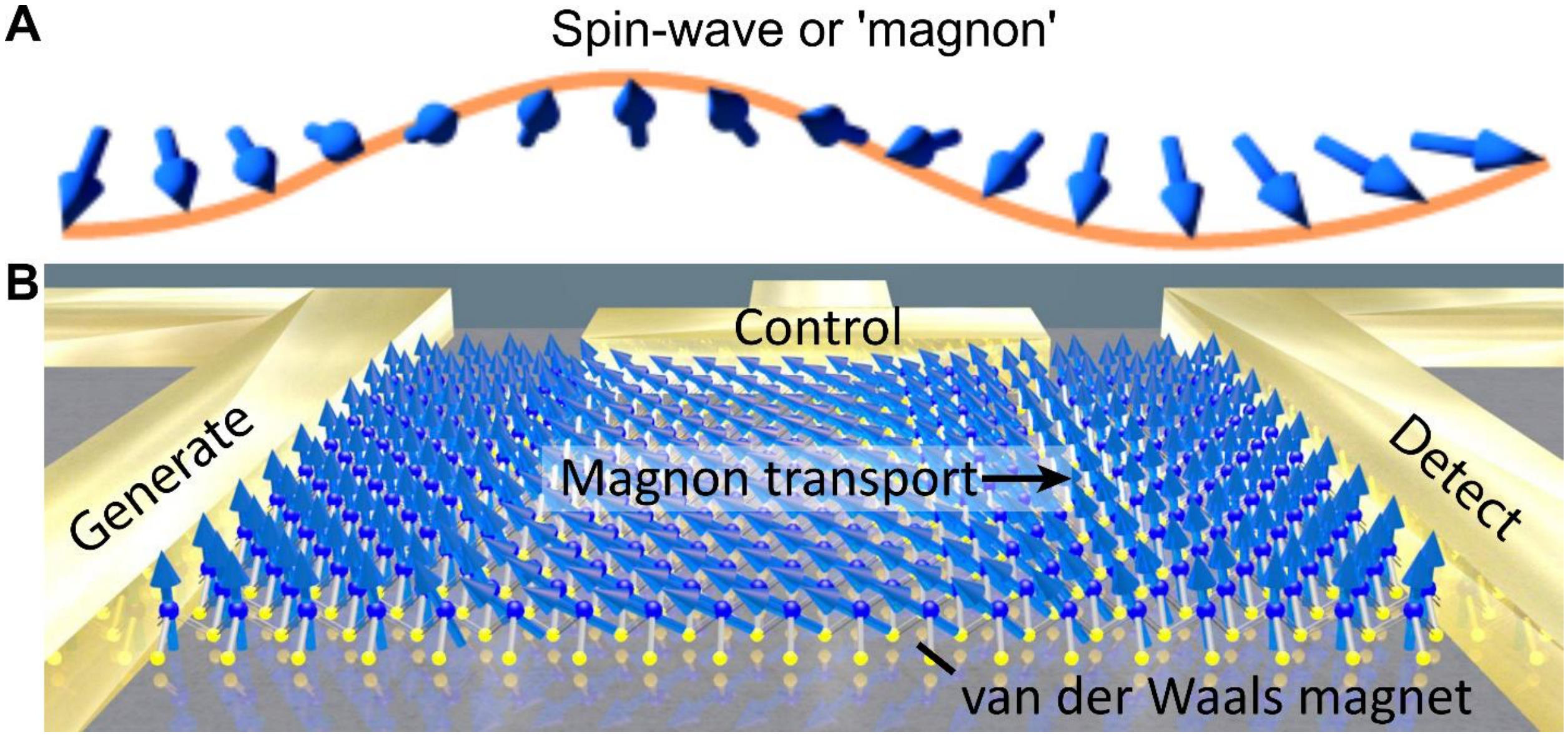

**Figure 1. Magnonics based on Van der Waals magnets.**
A) Illustration of a magnon/spin wave as an excitation of a homogeneous ferromagnetic state, where the spins are represented as blue arrows and the collective oscillatory behavior is highlighted as an orange trace. B) Magnon spintronics relies on manipulation and control of magnon spin transport from an injector to a detector. The electrical contacts for generating, controlling and detecting the magnons are shown as goldish pads, whereas the Van der Waals magnet is illustrated by blue and yellow balls (corresponding to magnetic and non-magnetic atoms, respectively). The spin is represented as a blue arrow, which orientation exhibits a spatial dependence that corresponds to the magnon being transported

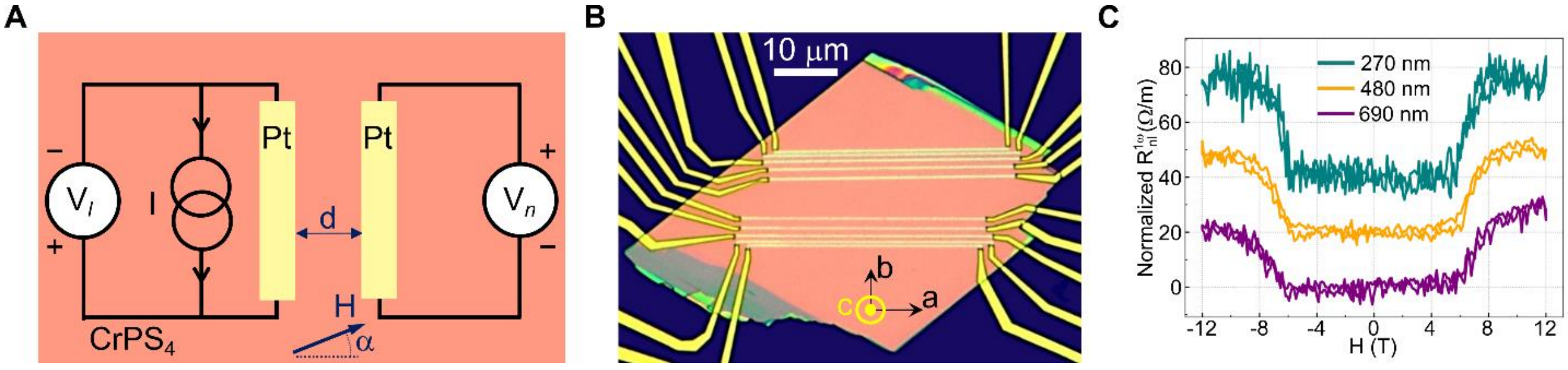

**Figure 2. Non-local magnon detection in Van der Waals magnets.**
A) Measurement configuration for local and non-local measurements of thermally and electrically injected magnon transport. The electrical circuit elements are represented as I, for the applied current, and $V_l$ and $V_{nl}$, for the measured voltages at the injector and detector, respectively. The injector and detector are platinum (Pt) strips separated by a distance *d*. An external applied magnetic field, *H*, is applied forming an angle *α* with respect to the normal direction of the Pt contacts (dashed line). B) Optical micrograph of Pt strips deposited on a $CrPS_4$ flake, indicating the crystal axes. C) Non-local measurements of electrically injected and detected magnon transport. The magnon transport arises above the spin-flip field of about 6 T. The traces correspond to different distances, d, between the platinum injector-detector contacts. Figure adapted from [40].